\documentstyle[buckow,epsf]{article}

\def\beq{\begin{equation}}
\def\eeq{\end{equation}}

\begin {document}

\def\titleline{
\vspace*{-60pt}
{\normalsize \hfill {\sf UTHEP-453}} \\
{\normalsize \hfill {\sf UTCCP-P-119}} \\
\vspace*{-2pt}
{\normalsize \hfill {\sf November, 2001}} \\
\vspace*{40pt}
Elementary Particles on a Dedicated Parallel Computer
%\newtitleline
}

\fnsymbol{footnote}
\setcounter{footnote}{1}

\def\authors{
Kazuyuki Kanaya\1ad for the CP-PACS Collaboration%
\footnote{
CP-PACS Collaboration:
        A.\ Ali Khan, 	
        S.\ Aoki,	
        G.\ Boyd,	
        R.\ Burkhalter,	
        S.\ Ejiri, 	
        M.\ Fukugita,	
        S.\ Hashimoto,	
	K.\ Ide, 	
        K.-I.\ Ishikawa,
        N.\ Ishizuka,	
        Y.\ Iwasaki,	
	T.\ Izubuchi, 	
        K.\ Kanaya,	
        T.\ Kaneko,	
        Y.\ Kuramashi,	
        V.\ Lesk, 	
        T.\ Manke, 	
	Y.\ Namekawa,	
        K.-I.\ Nagai, 	
	J.-I.\ Noaki, 	
        M.\ Okawa,	
	M.\ Okamoto,	
        Y.\ Taniguchi,	
        H.P.\ Shanahan, 
        A.\ Ukawa,	
        T.\ Umeda, 	
	T.\ Yamazaki,	
and     T.\ Yoshi\'e

\noindent
---------------------

Talk presented at the 18th AvH symposium
``100 Years Werner Heisenberg - Works and Impact'',
Bamberg, Germany, September 26-30, 2001.
}
}

\def\addresses{
\1ad
Institute of Physics, University of Tsukuba, 
Tsukuba, Ibaraki  305-8571, Japan\\
{\tt kanaya@rccp.tsukuba.ac.jp}
}

\def\abstracttext{
Numerical simulation of lattice QCD is now an indispensable tool for 
investigating non-perturbative properties of quarks at low energies. 
In this paper, I introduce a recent lattice QCD calculation of light 
hadron spectrum and quark masses 
performed on a dedicated parallel computer, CP-PACS. 
From a simulation of QCD with two flavors of dynamical quarks, we find 
$m_{ud}^{\overline{\rm MS}}(2{\rm GeV}) = 3.44^{+0.14}_{-0.22}$ MeV 
for the averaged mass of $u$ and $d$ quarks 
to reproduce the experimental masses of $\pi$ and $\rho$ mesons, and 
$m_{s}^{\overline{\rm MS}}(2{\rm GeV}) = 88^{+4}_{-6}$ MeV 
or $90^{+5}_{-11}$ MeV to reproduce the $K$ or $\phi$ meson mass. 
These values are about 20--30\% smaller than the previous estimates
using the quenched approximation.
}

\large
\makefront

%%%%%%%%%%%%%%%%%%%%%%
\section{Introduction}

Quarks are among the most fundamental building blocks of nature.
Unlike the previous elementary constituents, 
quarks are found to possess a remarkable property, {\it confinement}, 
{\it i.e.}, we cannot isolate quarks by dissociating hadrons.
Historically, fundamental theories of nature have emerged through 
reactions of isolated elements to weak external perturbations.
Quark confinement, however, has the consequence that we cannot directly 
measure the fundamental properties of quarks by experiment.

Fortunately, quarks are found to possess another remarkable 
property, {\it asymptotic freedom}, 
whereby interactions among quarks become weak at high energies.
This enabled us to identify quantum chromodynamics (QCD) as 
the fundamental theory of quarks.
QCD has been quite successful in explaining high energy processes 
of particles.

On the other hand, 
because the coupling parameter of QCD becomes large 
at low energies, basic properties of hadrons, such as mass spectrum, 
spatial sizes, decay constants, etc., are not calculable by conventional 
analytic methods. 
In other words, we cannot reconstruct hadrons from quarks based on 
the dynamics of QCD yet.
In order to answer the fundamental question of whether QCD is correct 
also at low energies, we have to carry out non-perturbative calculations
of hadrons directly from the first principles of QCD. 

Calculation of low-energy hadron properties is also required 
for determination of fundamental parameters of nature. 
Due to quark confinement, quark masses and coupling parameters
have to be inferred indirectly from a comparison of experimental results 
for hadron masses etc.\ with a theoretical calculation of them as 
functions of the fundamental parameters of QCD. 
Also for electroweak interactions, although the core part of 
the reaction can be reliably calculated in perturbation theory, 
uncertainties in QCD corrections to the quark currents propagate to 
the determination of the Cabibbo-Kobayashi-Maskawa 
parameters. 

Numerical simulation of quarks based on the lattice formulation of 
QCD is currently the only method to calculate low-energy
non-perturbative properties of hadrons. 
Recently, predictions with big impact are beginning to be produced 
through development of dedicated parallel computers \cite{dedicated}. 
In particular, a big breakthrough was achieved by the CP-PACS computer 
developed at the University of Tsukuba \cite{CPPACS}.
In this paper, I summarize major results from CP-PACS, 
focusing on the most fundamental topics of QCD, 
{\it i.e.}, the light hadron spectrum and light quark masses. 

In Sec.~\ref{sec:LQCD}, I introduce the lattice formulation of QCD
and explain why dedicated computers have been developed by physicists. 
Results for the light hadron spectrum and light quark mass are 
presented in Secs.~\ref{sec:Mhad} and \ref{sec:Mq}. 
Conclusions are given in Sec.~\ref{sec:conclusions}.

%%%%%%%%%%%%%%%%%%%%%%
\section{Lattice QCD and dedicated parallel computers}
\label{sec:LQCD}

We formulate QCD on a four-dimensional hypercubic lattice 
with a finite lattice spacing $a$ and finite lattice size 
$Na$.
The real world is defined by the limit of vanishing lattice 
spacing $a \rightarrow 0$ 
keeping the lattice size $Na$ finite or sufficiently large
(the continuum limit).
Before taking this limit, the theory is finite and mathematically 
well-defined.
Therefore, we can apply various non-perturbative techniques. 
Here, we perform numerical simulations on finite lattices
to calculate hadrons.

To accomplish this, 
the most time-consuming part of the calculation is the inversion of 
quark propagation kernels. 
Numerically, this is an inversion of large sparse complex matrices.
For the case of Wilson-type lattice quark actions, 
the typical size of the matrix is $12V \times 12V$, 
where $12=3\times4$ is the freedom of color and spin, and $V = N^4$ 
is the lattice volume. 
Because the condition number is inversely proportional to the quark mass, 
the inversion is numerically more intensive when we decrease 
the quark mass.
Even with the latest supercomputers, it is difficult 
to simulate the light $u,d$ quarks directly.
Therefore, in addition to the continuum extrapolation, 
we need to extrapolate the results at typically around the $s$ quark mass 
to the physical $u,d$ quark mass point. 
This procedure is called ``the chiral extrapolation''. 

In summary, in order to extract a prediction for the real world, 
we have to perform continuum and chiral extrapolations.
To get a precise and reliable result,
it is essential to have good control of these extrapolations. 
This requires a large-scale systematic calculation, 
and thus huge computer power.
%
%Accordingly, the birth of lattice QCD was possible first through 
%the availability of vector supercomputers in 1980's. 
%However, even for a simulation of single static hadron, 
The power of vector supercomputers in 1980's was not sufficient,
%Together with the rapid development of microprocessors in 1980's, 
which motivated several groups of lattice physicists to construct 
parallel computers dedicated to lattice calculations \cite{dedicated}.

\begin{figure}[t]

\begin{minipage}[t]{7.9cm}
\centerline{\epsfxsize=7.9cm \epsfbox{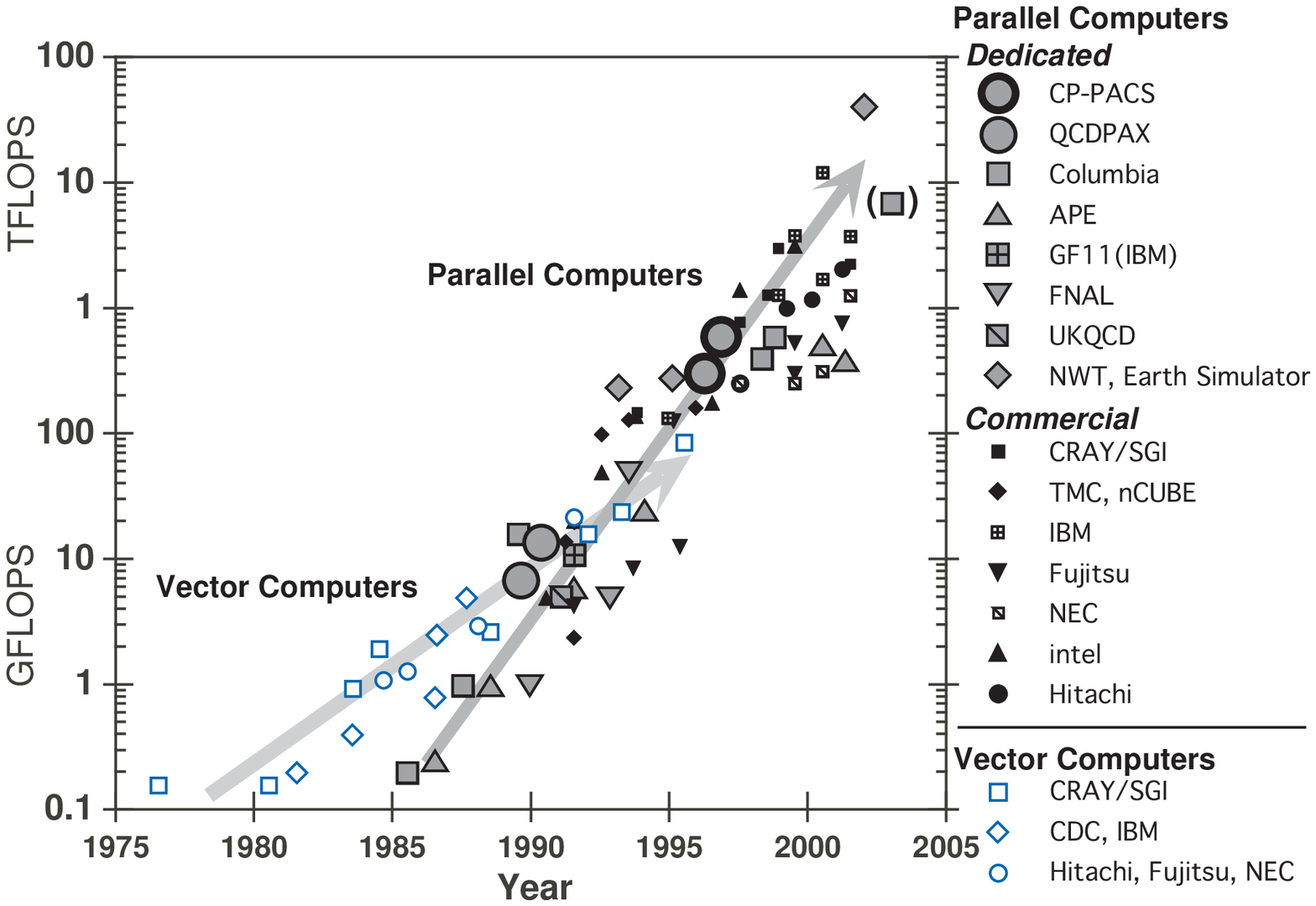}}
\vspace{-0.3cm}
\caption{Development of the computer speed.}
\label{fig:speed}
\end{minipage}
\hfill
\begin{minipage}[t]{7.8cm}
\centerline{\epsfxsize=6.3cm \epsfbox{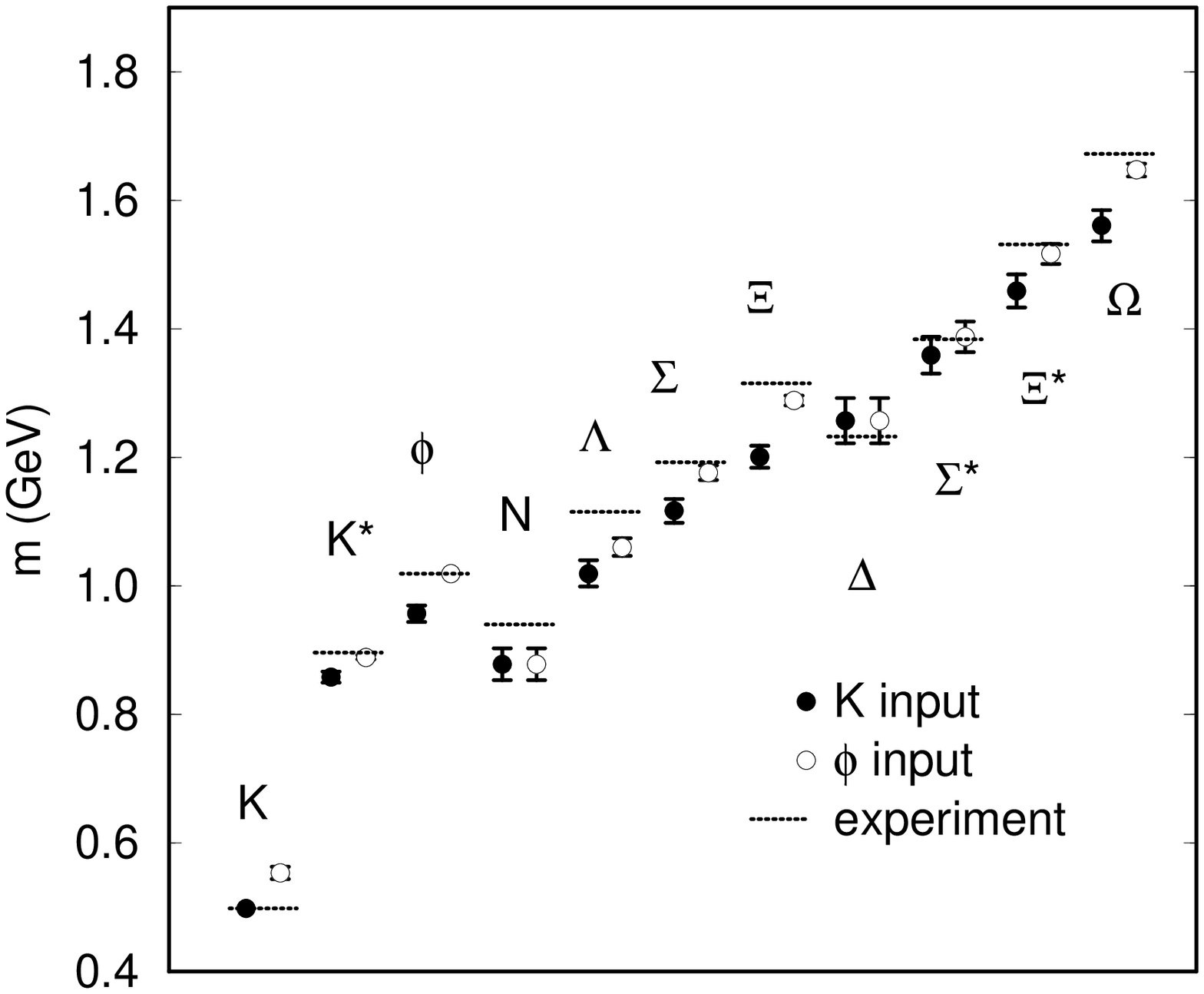}}
\vspace{-0.3cm}
\caption{Quenched light hadron spectrum \protect\cite{CPPACSquench}.}
\label{fig:quench}
\end{minipage}

%\vspace{-0.5cm}
\end{figure}

The historical development of computer speed is shown in 
Fig.~\ref{fig:speed}.
The frontier of speed has been advanced by parallel computers 
since around 1990. 
Dedicated machines developed by physicists contributed much to this trend.
At the University of Tsukuba, we have developed two machines, 
QCDPAX (1990) \cite{QCDPAX} and CP-PACS (1996) \cite{CPPACS}.
CP-PACS is a parallel computer achieving peak performance of 
614.4 GFLOPS with 2048 single-processor nodes. 
% interconnected by a three-dimensional hypercrossbar network. 
Since the first power-on in 1996, intensive lattice QCD calculations 
have been made on CP-PACS. 
In the following, I focus on the studies of hadron spectrum and 
light quark masses.

%%%%%%%%%%%%%%%%%%%%%%
\section{Light hadron mass spectrum}
\label{sec:Mhad}

Precise calculation of the hadron mass spectrum directly from 
the first principles of QCD is one of the main goals of lattice QCD. 
Because the computer power required is enormous, 
we study the issue step by step as follows:
\begin{description}
\item[Step 1]:
Calculate in the quenched approximation, in which the effects of 
dynamical pair creation and annihilation of quarks are neglected.
\item[Step 2]:
Include the dynamical $u,d$ quarks in a degenerate approximation, 
while heavier quarks are treated in the quenched approximation 
({\it two-flavor full QCD}).
\item[Step 3]:
Include the dynamical $s$ quark
({\it 2+1 flavor full QCD}).
\item[Steps 4, 5, $\cdots$]:
Introduce the $u,d$ mass difference, dynamical $c$ quark, etc.
\end{description}

\subsection{Quenched studies}
\label{sec:qMhad}

With the quenched approximation, we can reduce the computer time by a 
factor of several hundred preserving the basic properties of QCD 
(confinement, asymptotic freedom, and spontaneous breakdown of chiral
symmetry). 
The effects of the approximation are expected to be about 10\% 
in the spectrum. 

The first studies in this approximation had already been made in early 80's.
The issue turned out to be quite tough and computationally demanding. 
Actually, it took about ten years until the first systematic study, 
performing all the extrapolations, was attempted in 1993 \cite{GF11}.
From this study, the quenched light hadron spectrum was found to be 
consistent with experiment within the errors of about 10\%. 
However, the quality of data was insufficient to test the accuracy 
of numerical extrapolations, 
and also the final errors were too large to resolve the quenching artifact.

The status was significantly improved by CP-PACS 
\cite{CPPACSquench}. 
From an intensive computation involving about 100 times more 
floating-point calculations than the previous studies, 
it became possible to perform systematic tests on the quality of 
the extrapolations. 
Errors in the final results for the light hadron spectrum are now 
confidently estimated to be less than about 1\% for mesons and 
about 3\% for baryons. 
These errors include statistical and all systematic errors
except for those from the quenched approximation itself.

The quenched light hadron spectrum is shown in Fig.~\ref{fig:quench}.
Experimental values of meson masses --- $M_\pi$, $M_\rho$, and either 
$M_K$ or $M_\phi$ --- are used as inputs to fix the lattice spacing $a$ 
(or the strong coupling constant), 
the average $u,d$ quark mass $m_{ud}$, and the $s$ quark mass $m_s$. 
Masses for other hadrons are among the predictions of QCD. 
%Experimental masses are shown by horizontal bars in Fig.~\ref{fig:quench}.
%
From Fig.~\ref{fig:quench}, we find that the global pattern 
of the spectrum is well reproduced. 
At the same time, 
we clearly see that the quenched spectrum deviates from experiment 
by about 10\% (7 standard deviations) for the worst case:
Results from the $M_K$-input and $M_\phi$-input are discrepant by, 
and the hyperfine splitting between $K$ and $K^*$ mesons is smaller than 
experiment by about 10\%.
Decuplet baryon splittings are also small. 
Because all other systematic errors are well controlled, 
we identify the discrepancy as quenching artifact. 

\begin{figure}[t]

\begin{minipage}[t]{7.8cm}
\centerline{\epsfxsize=6.4cm \epsfbox{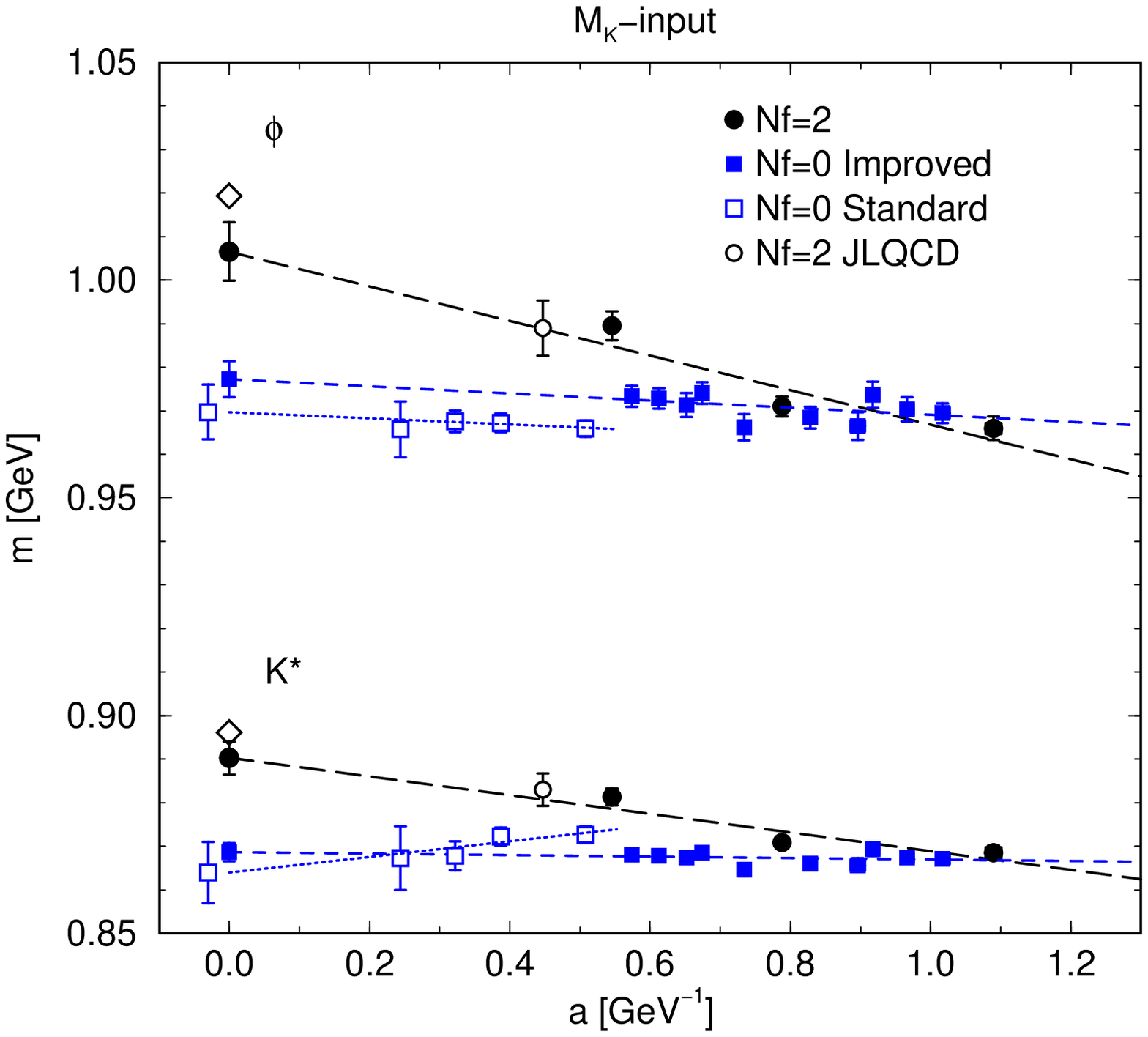}}
\vspace{-0.3cm}
\caption{Continuum extrapolation of vector meson masses $M_{\phi}$ and
$M_{K^*}$ in two-flavor full QCD ($N_f=2$) and quenched QCD ($N_f=0$), 
using $M_K$ as input \protect\cite{CPPACSfull}.
Open circles are recent results %of the JLQCD Collaboration 
using a different improved lattice action \protect\cite{JLQCD}.}
\label{fig:fspectrum}
\end{minipage}
\hfill
\begin{minipage}[t]{7.8cm}
\centerline{\epsfxsize=7.8cm \epsfbox{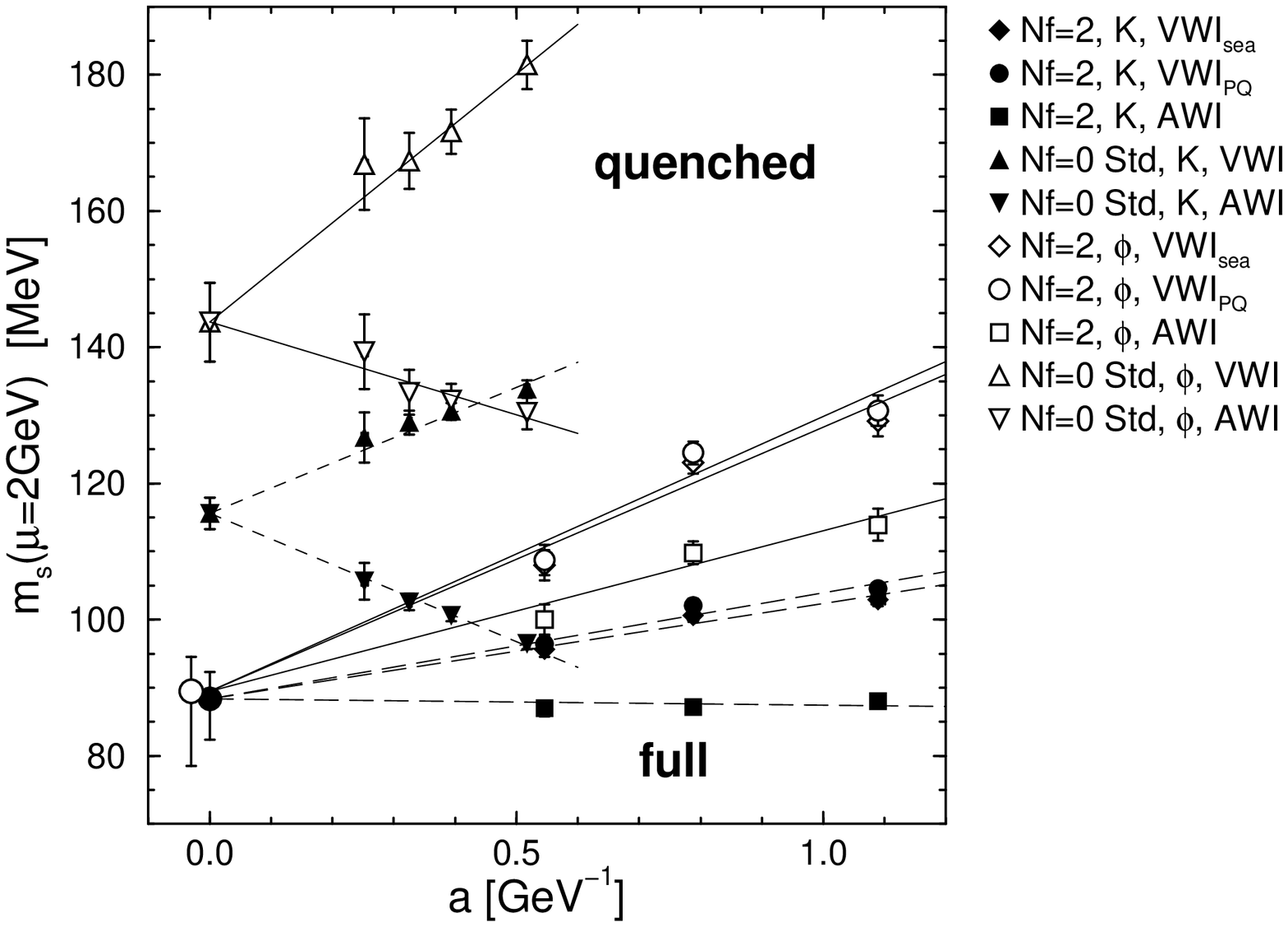}}
\vspace{-0.3cm}
\caption{Continuum extrapolation of the $s$ quark mass $m_s$ 
in the $\overline{\rm MS}$ scheme at 2 GeV
from two-flavor  \protect\cite{CPPACSfull} 
and quenched QCD \protect\cite{CPPACSquench}.
Quenched masses are shown by triangles. 
Filled and open symbols are for the $M_K$ and $M_\phi$ inputs,
respectively.
}
\label{fig:ms}
\end{minipage}

%\vspace{-0.2cm}
\end{figure}

\subsection{Two-flavor full QCD}
\label{sec:fMhad}

From the quenched simulation we find that, 
to calculate hadronic quantities with precision better than 10\%, 
we have to incorporate dynamical quarks. 
A naive extension of the quenched simulation to full QCD is difficult
because several hundred times more computer time is required. 
A partial solution is given by improvement of the 
lattice theory, whereby continuum properties are realized on 
coarser lattices. 
From a preparatory study \cite{comparative}, we find that the combination
of renormalization-group improved glue action and clover-improved
Wilson quark action is effective in removing major lattice artifacts. 
This reduces the computer time by a factor of about ten.

Although more might still be hoped for, we can start the first 
systematic studies of two-flavor full QCD on the CP-PACS 
performing both chiral and continuum extrapolations
\cite{CPPACSfull}. 
%For completeness, 
To identify dynamical quark effects clearly, 
we carried out another quenched simulation using the same improved 
action. 

Figure~\ref{fig:fspectrum} shows the continuum 
extrapolation of vector meson masses. 
We find that the two quenched results (open and filled squares) 
lead to universal values in the continuum limit $a=0$. 
They deviate from the experimental values (diamonds) as noted in the 
previous subsection. 
On the other hand, the full QCD results (circles) extrapolate to 
values much closer to experiment. 
Accordingly, we find no big discrepancies between the results from 
$M_K$ and $M_\phi$ inputs.
This means that the quenching artifacts in the spectrum are mostly 
removed by introducing dynamical $u,d$ quarks. 

The whole light hadron spectrum can now be approximately reproduced 
from QCD by adjusting just three parameters; the strong coupling constant, 
$m_{ud}$ and $m_s$. 
This provides us with a strong confirmation that QCD is correct also 
at low energies. 
%
%At the same time, our study demonstrates the importance of dynamical
%quarks in lattice QCD simulations.
%
Remaining small deviations from experiment may be explained by 
the quenched approximation of the $s$ quark. 
In order to confirm this, however, uncertainties from the chiral and 
continuum extrapolations should be reduced to the level of our quenched study.
This goal is reserved for future work.

\begin{table*}[t]
\caption{Light quark masses in MeV in the $\overline{\rm MS}$
scheme at 2 GeV \protect\cite{CPPACSfull}.}
\centerline{
\begin{tabular}{lcccc}
\hline
    &  $m_{ud}$ & $m_s$ ($K$-input) & 
       $m_s$ ($\phi$-input) & $m_s/m_{ud}$ \\
\hline
$N_f=0$ Stand. & 4.57$\pm 0.18$         & 116$\pm3$       & 144$\pm6$ 
& $\approx 25$--31 \\
$N_f=0$ Impr.  & 4.36$^{+0.14}_{-0.17}$ & 110$^{+3}_{-4}$ & 132$^{+4}_{-6}$ 
& $\approx 25$--30 \\
$N_f=2$ & 3.44$^{+0.14}_{-0.22}$ &  88$^{+4}_{-6}$ & 90$^{+5}_{-11}$ 
& 26$\pm2$\\
\hline
\end{tabular}
}
\label{tab:quarkmass}
\vspace{-0.2cm}
\end{table*}

%%%%%%%%%%%%%%%%%%%%%%
\section{Light quark masses}
\label{sec:Mq}

Adjustment of $m_{ud}$ and $m_s$ in the calculation of the hadron spectrum 
provides us with the most direct determination of quark masses from QCD.
Fig.~\ref{fig:ms} shows $m_s$ 
in the $\overline{\rm MS}$ scheme at $\mu=2$ GeV 
%from two-flavor QCD and quenched QCD 
as functions of the lattice spacing $a$. 
Results from the quenched improved action, which are consistent with 
the quenched standard action in the continuum limit, are omitted for 
clarity.
%The quark masses are converted into 
%the $\overline{\rm MS}$ scheme at $\mu=2$ GeV.

On the lattice, there are several alternative definitions for the quark 
mass, with the difference being $O(a)$.
In Fig.~\ref{fig:ms}, they are denoted as VWI 
%(vector Ward-Takahashi identity), 
and AWI (vector and axial-vector Ward-Takahashi identity). 
See \cite{CPPACSquench,CPPACSfull} for details. 
As shown in Fig.~\ref{fig:ms}, different definitions lead to different 
values of $m_q$ at finite lattice spacings. 
This has been a big source of uncertainty in the previous calculations. 
From Fig.~\ref{fig:ms}, we see that they converge to a universal value 
in the continuum limit, as theoretically expected. 
This confirms the quality of our calculations. 

The quenched value for $m_s$, however, differs by about 20\% between 
$M_K$-input and $M_\phi$-input. 
This is equivalent to the quenched artifact discussed in Sec.~\ref{sec:qMhad}. 
We note that the quenched artifact is larger in the light quark masses
than in the hadron spectrum. 
When we turn on the dynamical $u,d$ quarks, 
the discrepancy between the inputs disappears within our errors, 
in accord with the fact that the full QCD reproduces the 
hadron spectrum better (Sec.~\ref{sec:fMhad}).

The light quark masses in the continuum limit are summarized 
in Table~\ref{tab:quarkmass}. 
The errors include our estimates for systematic errors 
from chiral and continuum extrapolations and renormalization factors.
We note that the masses from two-flavor QCD are 20--30\% smaller than 
those from the quenched QCD. 
In particular, our $s$ quark mass in $N_f=2$ QCD is about 90 MeV, 
which is significantly smaller than the value $\approx 
150$ MeV often used in phenomenology. 
Our results are, however, consistent with recent estimates 
$m_s = 83$--130 MeV and $m_{ud} = 3.4$--5.3 MeV 
from QCD sum rules \cite{ChPT-SumRules} 
and $m_s/m_{ud}= 24.4\pm1.5$ from one-loop chiral perturbation 
theory \cite{ChPT-Quarkmass}.

%%%%%%%%%%%%%%%%%%%%%%
\section{Conclusions}
\label{sec:conclusions}

An intensive calculation of the light hadron spectrum in 
the quenched approximation has revealed discrepancies 
of about 10\% from experiment. 
With two flavors of dynamical $u,d$ quarks, 
the first systematic study performing both the continuum and 
chiral extrapolations has shown that the discrepancies are 
mostly removed, providing us with a strong confirmation 
that QCD is the correct theory of quarks at low energies as well. 

Precision calculation of hadron masses enables us to determine
fundamental parameters of quarks directly from the first principles 
of QCD.
We find that the dynamical quark effect is as large as 20--30\% 
in light quark masses. 
Noticeable and sizable dynamical quark effects are observed also 
in B meson decay constants \cite{CPPACSfB}, 
the equation of state at high temperatures \cite{CPPACSeos}, 
and the topological structure of the QCD vacuum \cite{CPPACSu1}. 
At the same time, new types of lattice fermions have begun to be 
applied to study hadronic matrix elements relevant to the 
$\Delta I =1/2$ rule and the CP violation parameters 
\cite{CPPACSdw}.

Because a shift in fundamental parameters can have 
significant implications to phenomenological studies of the 
standard model, 
it is urgent to evaluate dynamical quark effects in other 
hadronic quantities as well. 
The influence of dynamical $s$ quark should also be studied. 
% (Step 3).
Studies in these directions are under way 
to clarify the precise structure of the standard model. 

The studies presented in this report have been performed by the 
CP-PACS Collaboration. 
I thank the members of the Collaboration for discussions and 
comments.
This work is in part supported by 
the Grants-in-Aid of Ministry of Education, Science and Culture 
(Nos.~12304011 and 13640260) and JSPS Research for Future Program.

%%%%%%%%%%%%%%%%%%%%%%

\end{document}